# Starlink : A Solution to the Digital Connectivity Divide in Education in the Global South

H.M.V.R.Herath

Department of Electrical and Electronic Engineering, University of Peradeniya, Sri Lanka

**ABSTRACT**

Digital connectivity gap in the global south hampered the education of millions of school children during the COVID-19 pandemic. If not actions are taken to remedy this problem, future prospects of millions of children around will be bleak. This paper explores the feasibility of using the SpaceX Starlink satellite constellation as a means to alleviate the problem of the digital connectivity divide in the global south. First, the paper discusses the issues of digital connectivity in education in rural Sri Lanka and other countries in the global south. Then, the paper gives an introduction to Starlink broadband internet technology and discusses its advantages over traditional technologies. After that, the paper discusses a possible mechanism of adopting Starlink technology as a solution to the rural digital connectivity problem in the global south. Technological, as well as economical aspects of such scheme, are discussed. Finally, challenges that may arise in deploying a system such as Starlink to improve rural digital connectivity in Sri Lanka or any another country in the global south will be discussed with possible remedies.

I. INTRODUCTION

Online teaching/learning paradigm adopted all around the world as a consequence of SARS – Cov-19 pandemic put a sharp focus on the digital connectivity or lack thereof in the rural regions of the Global South including in Sri Lanka [1-4]. Primary, Secondary, and Tertiary education systems of the countries around the world had to adapt to an online delivery mode in haste due to pandemic imposed movement restrictions that made face-to-face learning/ teaching impossible. In Sri Lanka the mechanisms of delivery varied from the synchronous delivery of lessons using video conferencing technologies to sharing lessons and assignments through social media platforms. This is mainly true for the other countries of the global south as well. As a result of this diversity, the quality of education received by students varied a lot with rural students witnessing the already poor quality of education they receive further eroding. This widened the rural-urban education quality gap further. Poor digital connectivity is a major contributor to the poor quality of the online education received by students in rural areas. The impact is hardest felt in the primary and secondary education where UNICEF estimated that more than 168 million children lost more than full year of school education.

Urban centers of Sri Lanka, for example, are provided with internet connectivity via fiber to the home (FTTH) and mobile broadband technologies. These technologies provide reasonably sufficient bandwidth to the subscribers [1]. But, in certain urban pockets, the connectivity is not reliable due to the insufficient capacity of base stations and shadowing effects in the case of mobile broadband. When it comes to rural areas fiber connectivity to homes is not available due to economic factors such as the cost of deployment and insufficient subscriber base due to low population density as well as the poverty among the rural population. Furthermore, mobile broadband connectivity is sketchy in rural areas due to the light deployment of base stations. As a result, certain locations do not have connectivity and in many locations where the connectivity is available, the poor signal quality makes the user experience below par. Economics of deployment discussed earlier prevents internet service providers from expanding their networks deep into rural areas. As a consequence of above factors internet connectivity in Sri Lanka is about 50% according to 2020 data [5].

One could argue that, as education is a basic human right, the access to the medium through which the education is carried out needs to be a basic human right as well. In that context, it is imperative that people who live in every region of a certain country should have reasonable digital connectivity thereby allowing them to participate in educational and economic activities through the internet on an equal footing to every other citizen. This paper discusses how the connectivity provided by the Startlink satellite system can be utilized to provide equitable digital connectivity across any country in the global south [6].

The paper is organized as follows. Section II gives an introduction to the Starlink technology and compares it with other technologies that provide digital connectivity. Some field results of Starlink connectivity performance will be discussed in section III. After that, in the section IV a framework to incorporate Starlink to improve rural digital connectivity will be discussed. Possible challenges to this scheme and potential solutions will be discussed in section V. Finally, conclusions summarize the content of the paper.

II.　The starlink technology

Sir Arthur C. Clarke's concept of worldwide radio coverage using three geostationary satellites presented in 1945 marks the beginning of the satellite age. Satellite technology has advanced many folds since that publication through various technology cycles. The concept of low latency high bandwidth communication via a network of medium earth orbit (MEO) and low earth orbit (LEO) satellites gathered momentum in the late 80s. Iridium, ICO, and Globalstar are such systems deployed in the late 90s. But with the rapid improvements in terrestrial optical fiber and cellular wireless communication systems, the necessity of such satellite networks diminished. As a result, those MEO/LEO systems failed financially. In the second decade of the 20th century, with the rapid advancements of the ICT and increased requirements for communication capacity, the importance of the LEO satellite constellations, with its low latency and high throughput, re-emerged [7]. The inability of optical fiber and wireless communication technologies to penetrate remote areas of the globe due to various reasons act as a catalyst to this development.

Out of this background, several LEO constellation concepts emerged such as Kuiper, Starlink, Telesat, and OneWeb. These systems are at various stages of their development and use advanced technologies for spectrum usage, satellite and constellation throughput, ground equipment development, and management of systems. Out of these constellations, Starlink is the most extensive and the one that progressed most towards the deployment. Therefore, Starlink can be considered as the most promising solution to overcome the digital connectivity divide in education at this moment in time [6 - 9]

Federal Communication Commission (FCC) of USA in 2018 has granted SpaceX Starlink permission to deploy a constellation of LEO satellites in five orbital shells. SpaceX Starlink has so far deployed more than 1500 satellites of the 1st shell at an orbit of 560 km and at 53.00 inclination [6,7,9-13]. Figure 1 shows the Starlink satellite map as of 21st May 2021 [10]. They are stationed on 72 orbital plains with approximately 20 satellites on each plain [13]. Satellites operate in Ku-band and each has a mass of approximately 240 kg [6]. The system uses phase array antennas for up and downlinks and laser communication in the inter-satellite link (ISL) [14,15]. Considering the fact that laser communication occurs in a vacuum, light travels 47% faster than in the terrestrial fiber network. A latency of less than 30 ms is expected in this network.

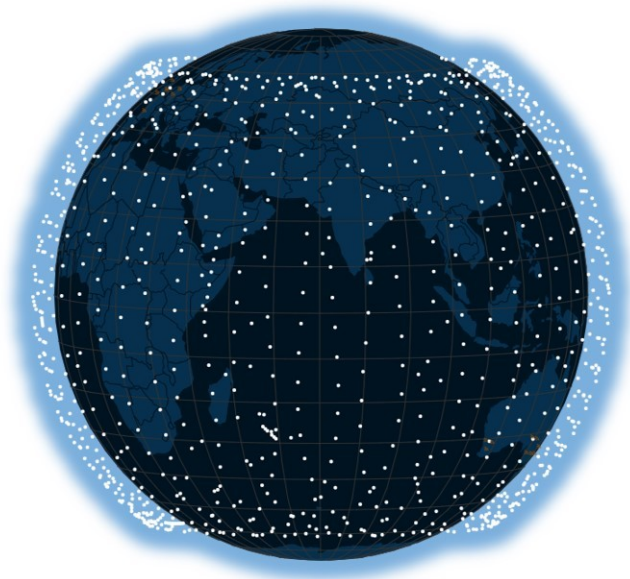

Figure 1: Starlink Satellite Map as of 21st May 2021 [10]

The architecture of the Starlink system is shown in Figure 2. Ground stations or Starlink Gateways are in constant communication with the satellites [9,14,15]. They provide internet access and control information to user terminals. User-Satellite communication uses Ku band and Ground Station-Satellite communication uses Ku band for downlink and Ka band for uplink [6, 16]. SpaceX's satellites generate ultra-small spot-size beams due to the fact that they are much closer to the earth compared to geostationary satellites. Close proximity to the earth provides higher speed and lower latency. The estimated total bandwidth throughput at the start of the commercial deployment is 23.7 Tbps [6].

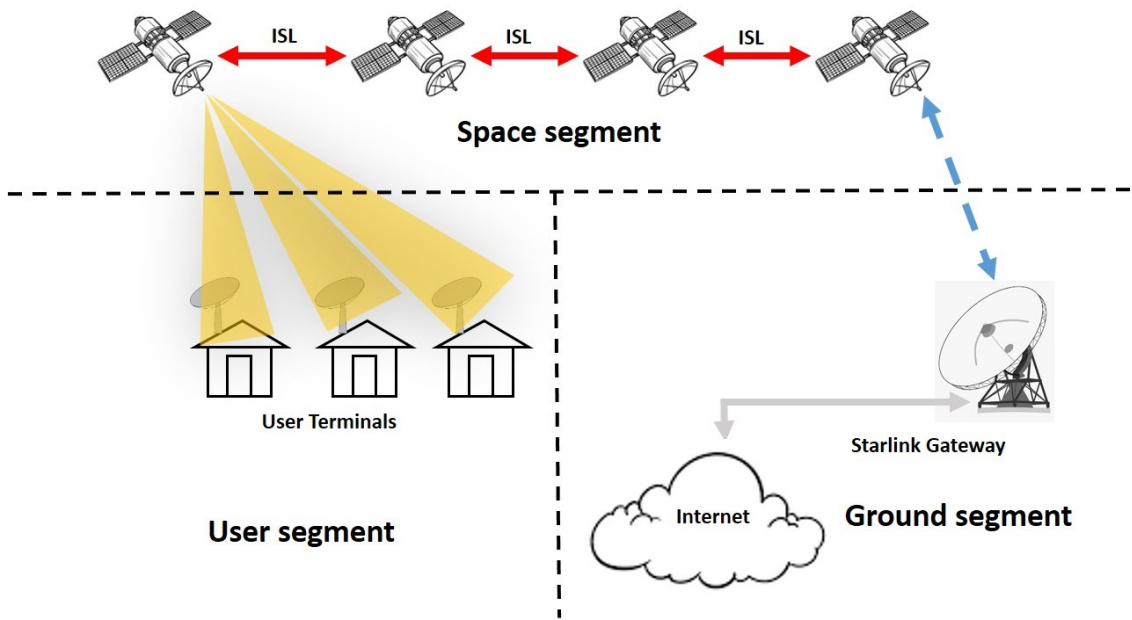

Figure 2: Startlink System Architecture

Starlink customer premises equipment (CPE) consists of a satellite dish, a Wi-Fi router, and a power supply unit. This is a plug and play system. The dish is 23" in diameter and can be easily handled by a single person [15,17]. It can be placed on the ground or on a rooftop where clear sky visibility is available. The dish consists of a phased antenna array of the stacked honeycomb structure. The dish can automatically align with the available Starlink satellite. Starlink uses advanced phased-array technology for both the satellite and the customer dish [15-17]. That allows nearly instantaneous hand-offs between different satellites with no mechanical transitions.

The router is equipped with a Gigabit Ethernet port and Wi-Fi to provide connectivity. The satellite dish is connected to the router and both are powered using Power over Ethernet (PoE). One router can support up to 128 devices simultaneously [18]. It is operated with a 56 V DC supply provided through PoE. The router complies with IEEE 802.11 standard and operates at 2.4 GHz and 5 GHz. OFDM modulation technology is used for transmission [15-17].

## III. STARLINK USER END PERFORMANCE

In October 2020 public beta trial program for subscribers in the northern United States and Canada between the latitudes of 45º and 52º was commenced. Starlink expected to provide full equatorial coverage by the beginning of 2022. Starlink services will be available in Sri Lanka in 2022 Subjected to the regulatory approval according to the Starlink web site [19]. At present Starlink provides unlimited internet access to the subscribers.

Independent performance analysis of the Starlink user terminal was carried out by the ROADMAP-5G research group of the Carintia University of Applied Sciences, Austria in June 2021[15]. Key findings of the experiment is shown in the Table 1. According to the experiment YouTube streaming provided satisfactory performance with a rare exception of having 4-6 second interruptions [15]. Automatic switching between the satellites follow a pre-defined timing of 15 seconds according to the observations [15]. Latencies fluctuated nearly always during this period [15].

The monthly subscription during the public beta is $99 for download speeds between 50–150 Mbps and latency between 20–40 milliseconds. A one-time equipment fee of $499 is also charged [19]. The data cap has not yet been implemented for the connection so far. Starlink expects to improve the download speed to 300 Mbps and reduce the equipment cost further according to the promotional materials.

The main advantage Starlink has over terrestrial cellular and fiber technologies is that it can be deployed in remote rural areas without much supporting infrastructure in an economical manner. In addition, Starlink can provide performance comparable or better than cellular systems in rural areas.

Table 1: Key performance parameters of Starlink [15]

| Parameter | Performance |
| --- | --- |
| Average download throughput | ~170 Mbps |
| Maximum download throughput | ~330 Mbps |
| Average upload throughput | ~17 Mbps |
| Maximum upload throughput | ~60 Mbps |
| Latency | 30ms – 2s |
| Percentage of time latency is below 90ms | 98% |
| Percentage of time latency is below 50ms | 77% |
| Down time | 2.4% |
| Average power consumption | 105 W |
| Peak power consumption | 190 W |

# IV. RURAL DIGITAL CONNECTIVITY VIA STARLINK

This section proposes a framework to provide digital connectivity to remote rural locations via Starlink. The proposal considers an off-grid location as an example. It can be easily modified for a grid-connected location. It is considered that 2 Mbps download speed and 0.25 Mbps upload speed is sufficient for a reasonable remote learning experience considering the fact that students are receiving content most of the time in a remote learning session [20]. Above mentioned numbers are viable according to the authors' own experience in delivering remote learning sessions.

Considering the average download throughput given in Table 1. 85 simultaneous connections can be supported by a single Starlink CPE. But, when the average upload speed is considered only 68 simultaneous connections can be supported by a single Starlink CPE. Allowing a tolerance for any performance fluctuations it is reasonable to assume that access to 50 devices can be given simultaneously through a single Starlink CPE using the current technology to provide a reasonable remote learning experience. With 50 simultaneous users, each user on average will have more than 3 Mbps download speed and more than 0.3 Mbps upload speed.

Figure 3 presents the conceptual idea of a rural off-grid digital connectivity center. In this center, internet connectivity is provided via Starlink and is powered by solar energy. First, let's do an approximate power requirement calculation. Assume that this center is equipped with 5 desktop computers and 5 laptops. Additionally, 40 tablets can be simultaneously used in this location. The physical structure of the center can either be constructed for the purpose or it may be a re-purposed existing building.

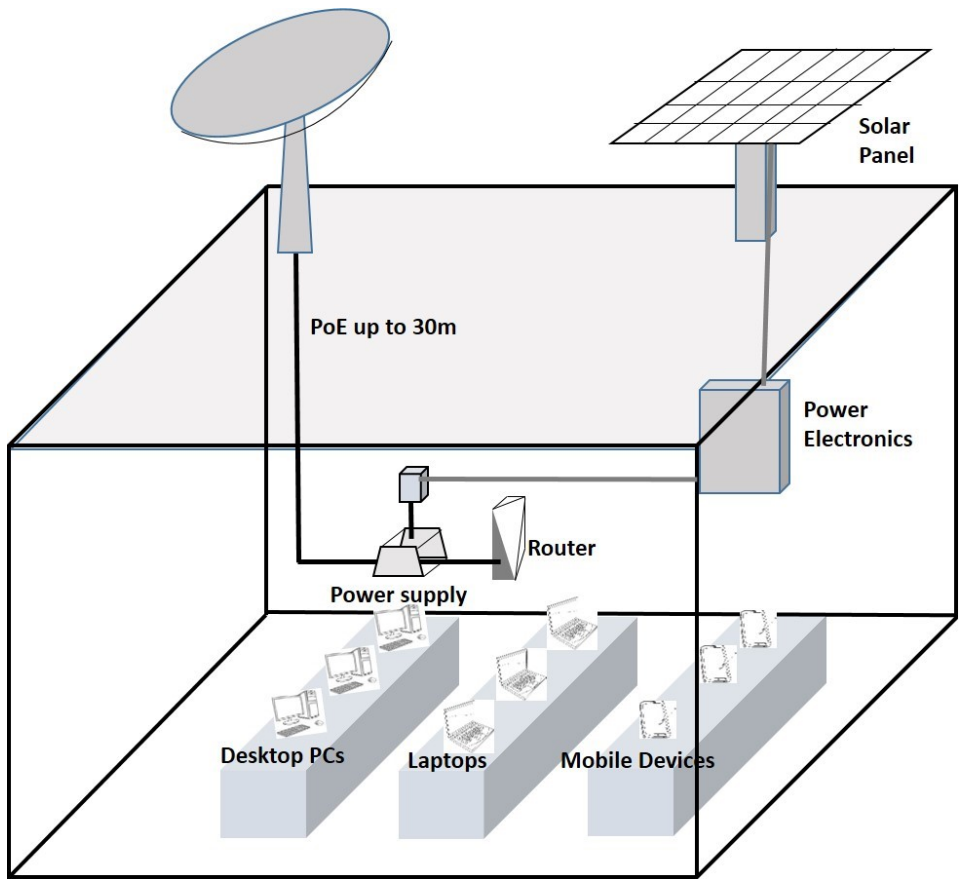

Figure 3: Proposed Rural Digital Connectivity Center

It is reasonable to assume that a desktop PC with an LED display consumes approximately 150 W of power and a laptop consumes 60W of power. Additionally, a tablet consumes approximately 10 W of power. So all the devices combine consume approximately 1.45 kW. Considering all the other requirements a 3 kW solar energy system would suffice to power up such a center. The capital cost of such a system would be 3500 USD equivalent Sri Lankan currency.

It would be useful to consider the economic viability of such a center with respect to the connectivity cost. Here, three possible utilization schemes are considered. In scheme 1, each user gets to use the center for a 4-hour time slot per day. Considering 8-hour daytime usage, 100 users can use the center per day. In scheme 2, each user gets to use the center for a 3-hour time slot per day. Considering 9-hour daytime usage, 150 users can use the center per day. In scheme 3, each user gets to use the center for a 2-hour time slot per day. Considering 8-hour daytime usage, 200 users can use the center per day. These number can be decided considering the ground situation such as number of students and the area of service.

Assuming that the same set of users use the center every day it is possible to calculate connectivity cost per user per month considering the Starlink subscription of 99 USD/month. It would be 99 cents, 66 cents, and 49 cents respectively for schemes 1, 2, and 3. The equipment cost per head would be one time 4.99 USD, 3.32 USD, and 2.49 USD expenditure respectively. Considering an equipment lifetime of 3 years, equipment cost would be negligible. In comparison two ISPs in Sri Lanka provide unlimited broadband access at up to 2 Mbps for a monthly rental of about 10 USD according to their promotional materials. Above calculations show that deploying rural digital connectivity centers linked by Starlink is an economically viable proposition for the countries in the global south. Considering the fact that services can be provided to 100-200 people per day justifies the capital expenditure to establish such a system. A Government institution, private corporations or NGOs could fund the establishment of the centers while the cost of the maintenance and subscription could be recovered from the users.

The primary objective of the digital connectivity center would be to enhance the educational opportunities of children via reliable digital connectivity. Additionally, resources could be used by adults for e-commerce in the evening and at night provided that there is sufficient energy storage. For a grid-connected digital connectivity center, a solar energy generation system is optional.

Considering the fact that this center is mainly used for educational activities a local school would be a best location to establish this center. Locations such as "e-nanasala", a digital connectivity providing center in Sri Lanka, or a community center could be alternatives to the schools to establish proposed center.

V.     CHALLENGES AND POSSIBLE SOLUTIONS

In order for Starlink to operate in any country, it is necessary to get regulatory approval to provide telecommunication services in that country. So far there is no evidence that the Starlink received such approval from any of the countries in the global south. It is expected that they will start that process soon.

Considering the Starlink network architecture, users could access internet without subjected to the control of governments of the respective countries. Governments could consider this as a threat to their sovereignty. Therefore, it is essential that Starlink come to a workable agreement with governments to solve this issue to the satisfaction of all parties concerned. One solution could be establishing at least one gateway in the country of interest and routing traffic through that gateway.

There will be a pushback from the established internet service providers (ISPs) within a country who might consider Starlink as a competitor. Furthermore, the quality of the broadband services in many countries in the global south are poor and it may be possible that Starlink would be able to provide better broadband service even in urban areas compared to the local ISPs. Therefore, it is essential that all the parties come together and find out ways of collaborating in such a way that all the parties including customers are benefitted. One possibility is using Starlink to improve the performance of cellular backhaul of the local service providers. Starlink would be able to connect user terminals to the local internet infrastructure through a gateway.

## VI. CONCLUSIONS

This paper puts forward a conceptual idea of how the Starlink satellite network could be utilized to provide digital connectivity to rural remote locations in order to overcome the digital connectivity divide in education in the global south. A proposal to establish rural digital connectivity centers is discussed in this paper. Through few simple calculations, it could be concluded that such a project would have long-term economic viability and could provide adequate capacity to service 50 users simultaneously from the existing technology. Possible regulatory and commercial challenges to a project of this nature were discussed with possible solutions.

Providing better educational opportunities to rural youth via reliable internet could move them out of poverty and allow them to give a better contribution to the local economy. Furthermore, the proposed center could promote e-commerce in rural areas.